\documentclass[twoside]{dis04}
\def\runauthor{Stefano Frixione}
\def\shorttitle{Bottom production}
\def\beq{\begin{equation}}
\def\eeq{\end{equation}}
\def\beqn{\begin{eqnarray}}
\def\eeqn{\end{eqnarray}}
\newcommand\sss{\scriptscriptstyle}

\newcommand\as{\alpha_{\sss S}}         
\newcommand\pt{p_{\sss T}}         
\newcommand\ptt{p_{\sss T}^2}
         
\newcommand\bb{\overline{b}}         
\newcommand\shat{\hat{s}}
\newcommand\logarg{{\cal Q}}
\newcommand\hatpt{\hat{p}_{\sss T}}
\newcommand\gsim{\mathop{\mbox{\vbox{\hbox{$>$} \vskip -9pt \hbox{$\sim$}
             \vskip -3pt  }}}}
\newcommand\lsim{\mathop{\mbox{\vbox{\hbox{$<$} \vskip -9pt \hbox{$\sim$}
             \vskip -3pt  }}}}

\begin{document}

\title{Bottom production}

\author{Stefano Frixione}

\address{INFN, Sezione di Genova\\ 
Via Dodecaneso 33, 16146 Genova, Italy\\
E-mail: Stefano.Frixione@cern.ch}

\maketitle

\abstracts{I briefly review the theory and phenomenology of bottom
production at colliders. When all theoretical uncertainties are taken
into proper account, and modern measurements are considered, no significant
discrepancy is seen between data and QCD predictions
}

\noindent
The physics of bottom quarks is one of the best-studied topics in
particle physics. Experimentally, this is due to the abundance with
which $b$ quarks are produced at colliders. Theoretically, the reasons
of interest are many. The characteristic that sets the $b$ apart is
its heaviness,
\beq
m_b\gg\Lambda{\sss QCD}
\label{bmass}
\eeq
which entails peculiar properties. If one is interested in the
phenomenology of the decays of the $b$-flavoured hadrons, eq.~(\ref{bmass})
suggests to treat the $b$ as infinitely heavy in comparison with its
companion light quark(s) in a bound state, paving the way to HQET and its
symmetry properties. On the other hand, if one aims at studying the
hard production mechanism, eq.~(\ref{bmass}) implies the possibility
of computing the open-$b$ cross section, which is free of collinear
and infrared singularities order by order in perturbation theory
(as opposed to, say, open-$u$ cross section, whose final-state
collinear singularities are cancelled only upon convolution with
a non-perturbative fragmentation function). The bottom is also the
heaviest quark which hadronizes before decaying, allowing us to 
test many of the ideas of the factorization theorems in a relatively
clean environment.

To be definite, let me consider the $b$ cross section in hadronic collisions,
which can be written as follows:
\beqn
d\sigma_{H_1H_2\to b\bb}(S)=\sum_{ij}\int dx_1 dx_2 
{f_i^{(H_1)}}(x_1){f_j^{(H_2)}}(x_2)
{d\hat{\sigma}_{ij\to b\bb}}(\shat=x_1 x_2 S),
\label{factth}
\eeqn
and can be readily extended to other types of colliding particles.
Here, $f_i^{(H)}$ are the parton distribution functions (PDFs), and
$d\hat{\sigma}_{ij\to b\bb}$ are the short-distance cross sections,
the only pieces in eq.~(\ref{factth}) that can be computed in perturbation
theory. The LO term (of ${\cal O}(\as^2)$) is trivial to obtain. The
NLO one (of ${\cal O}(\as^3)$) was the result of landmark 
calculations~\cite{Nason:1987xz,Nason:1989zy,Beenakker:1990ma}.
Not surprisingly, the NNLO term is not available at the moment; this
may be worrisome, since at the NLO the scale dependence is still pretty
large, and the corrections are 100\% of the Born term. However, there
are at least a couple of points that deserve more immediate attention
than the lack of the NNLO contribution. The first is that, as in any other 
cross section computed in perturbation theory, large logs can appear
which spoil the ``convergence'' of the series. In other words,
\beqn
d\hat{\sigma}=\sum_{i=2}^\infty a_i\as^i\,,\;\;\;\;\;\;
a_i=\sum_{k=0}^{i-2}a_i^{(i-2-k)}{\log^{i-2-k}\logarg}\,,
\label{FOser}
\eeqn
where $\logarg$ generically indicates a ``large'' quantity, 
in the sense that $\as\log^2\logarg\gsim 1$. The second problem is
that, although theoretically well defined, the open-$b$ cross section
is not physically observable. In order to compare theoretical predictions
with data, two things can be done. {\em a)} Hadron-level experimental
data are deconvoluted, and presented in terms of parton-level ``measurements''
that can be directly compared with open-$b$ results. These deconvolutions
are typically performed by means of parton shower Monte Carlos by the
experimental collaborations. {\em b)} The open-$b$ cross section is
convoluted with a non-perturbative fragmentation function (NPFF)
$D^{b\to H_b}$; for the single-inclusive $\pt$ spectrum, one writes
\beq
\frac{d\sigma(H_b)}{d\pt}=
\int\frac{dz}{z}{D^{b\to H_b}}(z,\epsilon)
\frac{d\hat{\sigma}(b)}{d\hatpt},
\phantom{aaaaaaaa}
\pt=z\hatpt\,.
\label{conv}
\eeq
$D^{b\to H_b}$ describes how a $b$ quark transforms (``fragments'') into
a $B$ hadron; it is not computable in perturbation theory but, being
universal in the same sense as PDFs, can be fitted to data in a given
type of collision (usually $e^+e^-$) and used elsewhere.

Common wisdom has it that neither strategy {\em a)} nor {\em b)} have
been particularly successful, since Tevatron data (and, to some extent,
SpS ones) have been shown to be systematically larger than NLO QCD
predictions, regardless of whether they were presented in terms of $b$ 
quarks or of $B$ mesons. In a recent CDF paper~\cite{Acosta:2001rz}
on $B^\pm$ single-inclusive $\pt$ spectrum,
the discrepancy was quantified to be $2.9\pm 0.2\pm 0.4$. Taking these
comparisons blindly, one is led to conclude that $b$ physics is {\em the}
problem of the SM, and offers the first glance beyond it~\cite{Berger:2000mp}.
Although this remains a viable possibility, it seems premature to buy it
without first reassessing carefully all possible sources of mistakes
in the past comparisons between theory and data, and considering the
uncertainties that so-far uncalculated SM contributions can give.
In particular, one should try to answer the following questions:

{\em 1)} Do large logs spoil the convergence of the series?

{\em 2)} Is the fragmentation/deconvolution performed appropriately?


\noindent
But before getting into this, let me point out that, although the
discrepancies between data and NLO QCD have been quoted to be large,
this is mainly due to the failure to incorporate properly {\em all} 
the uncertainties, including the theoretical ones which are very large. 
Upon doing so one realizes that, on a statistically sound basis, most of the
data lie withing $1\sigma$ from the default theoretical predictions, and very
rarely the discrepancy exceeds the $2\sigma$ level. The interested reader can
find an informative discussion in ref.~\cite{MLMtalk}. 

The answer to question {\em 1)} depends on the fact that the 
logarithms that grow potentially large can be
divided into two classes. The first class includes those logs whose
arguments don't depend on the observable being measured, such as
\beq
\logarg=1-\frac{4m_b^2}{\shat},\;\;\;\;
\logarg=\frac{m_b^2}{\shat},
\eeq
which are known as threshold logs (relevant when the c.m. energy is 
not much larger than the quark mass), and small-$x$ logs (relevant when 
the quark mass is negligible wrt the c.m. energy) respectively. Threshold
logs are clearly not a factor at colliders; small-$x$ logs are estimated
to give up to 30\% effects at the Tevatron~\cite{Collins:1991ty}. Thus, 
the overall picture would not change if these logs were properly resummed 
and matched with NLO QCD predictions. The logs belonging to the second 
class do have arguments which are directly related to the observables.
For example, when measuring the single-inclusive $\pt$ spectrum, the $b\bb$
$\pt$ spectrum, or the $b\bb$ azimuthal distance in the transverse plane, 
logs of the following arguments are generated
\beq
\logarg=\frac{\pt(b)}{m_b},\;\;\;\;
\logarg=\frac{\pt(b\bb)}{m_b},\;\;\;\;
\logarg=1-\frac{\Delta\phi(b\bb)}{\pi},
\eeq
and the perturbative series may be badly behaved when $\pt(b)\gg m_b$,
$\pt(b\bb)\simeq 0$, and $\Delta\phi(b\bb)\simeq \pi$ respectively.
Since single-inclusive $\pt$ spectra are routinely measured, large
$\pt(b)/m_b$ logs have a prominent role. The resummation of these logs, 
i.e. the rearrangement of the perturbative series of eq.~(\ref{FOser})
in the following form:
\beq
\frac{d\sigma}{d\ptt}=\as^2\sum_{j=0}^\infty\sum_{i=0}^\infty
r_i^{(j)}\as^j\left(\as\log\frac{\ptt}{m_b^2}\right)^i+{\rm PST}
\label{RESser}
\eeq
can be achieved by using the perturbative fragmentation functions
computed in ref.~\cite{Mele:1990cw}. Each term in the sum which runs
over $j$ corresponds to a given logarithmic accuracy ($j=0$ is LL,
$j=1$ is NLL, and so on), and PST stands for power suppressed terms,
i.e. terms which vanish in the limit $m/\pt\to 0$. Typically,
eq.~(\ref{RESser}) is computed up to the NLL; since PST are neglected,
the quark behaves as if massless, which implies that predictions based
on eq.~(\ref{RESser}) must strictly be used only when $\pt\gg m_b$.
The trouble is that $\pt\gg m_b$ is not a quantitative statement, and often
resummed predictions are compared to data even for $\pt\lsim m_b$.
It should be clear that such a comparison is void of sense, and any
agreement between theory and measurement must be regarded as accidental.
In any events, a matched computation (FONLL) was proposed in 
ref.~\cite{Cacciari:1998it}, which combines the virtues of the fixed
order and of the resummed formulae:
\beq
\frac{d\sigma}{d\ptt}=
a_2\as^2+a_3\as^3+
\as^2\sum_{i=2}^\infty {r_i^{(0)}}
\left(\as\log\frac{\ptt}{m_b^2}\right)^i+
\as^3\sum_{i=1}^\infty {r_i^{(1)}}
\left(\as\log\frac{\ptt}{m_b^2}\right)^i
\label{FONLL}
\eeq
and can be used to get sensible predictions in the whole $\pt$ range.
Armed with eq.~(\ref{FONLL}), one can answer question {\em 1)} in a
quantitative way; it turns out that, in the $\pt$ range probed at the Tevatron,
the effects are moderate. As discussed in ref.~\cite{Cacciari:2002pa},
the NLO cross section used in ref.~\cite{Acosta:2001rz} is only about 
20\% lower than the FONLL one.

Let me therefore consider question {\em 2)}, and for simplicity discuss
the case of fragmentation rather than that of deconvolution. The master
equation is~(\ref{conv}). The crucial point is that, while the l.h.s.
of this equation is a measurable quantity, neither of the two terms
on the r.h.s. is measurable. This is easy to understand if one considers
that eq.~(\ref{conv}) is used to extract the NPFF from data; the l.h.s.
is measured, the short-distance cross section $d\hat{\sigma}$ is
computed, and eq.~(\ref{conv}) is solved for NPFF:
symbolically, NPFF=$d\sigma(H_b)/d\hat{\sigma}(b)$. Thus,
if one uses $d\hat{\sigma}$ computed at the LO, the resulting NPFF
will clearly differ from the one obtained by computing $d\hat{\sigma}$
to, say, NLO. Notice that this doesn't contradict the universality 
property of the NPFF; this property merely states that a given NPFF
stays the same regardless of the type of hard collisions involved.
It follows that sensible predictions can be obtained only if the NPFF
has been extracted from $e^+e^-$ data using a cross section computed
in the same approximation as that used to predict the $p\bar{p}$ 
cross section. By reconsidering carefully the fragmentation procedure
adopted in ref.~\cite{Acosta:2001rz}, the authors of 
ref.~\cite{Cacciari:2002pa} pointed out that the claimed discrepancy
of $2.9\pm 0.2\pm 0.4$ turns actually out to be $1.7\pm 0.5\pm 0.5$
(this also includes the 20\% effect mentioned before, for replacing
the NLO result with the FONLL one), i.e. data
are within $1\sigma$ from the default theoretical prediction.
This finding is consistent with experimental evidence from 
D0~\cite{Abbott:2000iv} that the inclusive rate of jets containing 
$b$ quarks -- a quantity largely insensitive to the details of
the perturbative and non-perturbative fragmentation -- agrees with 
NLO QCD predictions~\cite{Frixione:1996nh}. 

\begin{figure}[t]
\begin{center}
 \epsfig{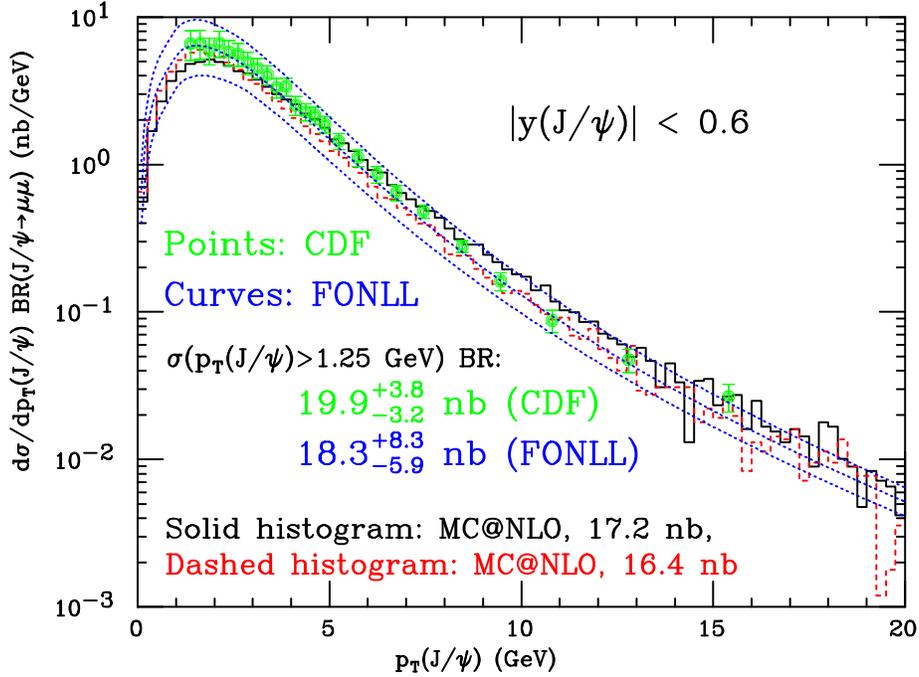}
\end{center}
\caption[*]{Comparison of CDF Run II data~\cite{cdfrun2} with
MC@NLO~\cite{Frixione:2002ik,Frixione:2003ei} and
FONLL~\cite{Cacciari:1998it} predictions.}
\label{fig:CDFrun2}
\end{figure}
More evidence that things go in the right direction has been achieved
thanks to a new CDF measurement~\cite{cdfrun2} of single-inclusive
$b$-hadron $\pt$ spectrum in the central rapidity region. For the first
time, these data probe the region of $\pt\simeq 0$, where fragmentation
effects play a minor role; thus, an improved agreement wrt the previous
comparisons would support the conclusion that what we have to blame
is our incomplete understanding of the fragmentation phase. This is
in fact what happens. The data show the best-ever agreement 
(see fig.~\ref{fig:CDFrun2}) with FONLL and 
MC@NLO~\cite{Frixione:2002ik,Frixione:2003ei}. I stress that FONLL
and MC@NLO are both based on the NLO computations of ref.~\cite{Nason:1987xz},
and they differ only in the resummation of logs beyond the leading ones,
and in the treatment of the $b$ quark hadronization and decay, which
in MC@NLO is performed through the HERWIG~\cite{Corcella:2000bw} cluster model.
It is remarkable that, in spite of the differences in the treatment of
the resummation of the large-$\pt$ logs, FONLL and MC@NLO can be made to
agree perfectly with a proper tuning of the fragmentation and of the
clustering parameters (see ref.~\cite{Cacciari:2003uh} for a discussion
on this point and on the comparison between theory and Run II CDF data).

Let me now turn to $b$ cross section measurements at HERA. A couple
of years ago, the situation appeared to be consistent with what was
observed at the Tevatron, with NLO QCD predictions systematically
undershooting H1 and ZEUS data. This picture has now radically
changed, as has been thoroughly documented at this 
conference~\cite{batHERA}. All data lie within $2\sigma$ from
the theoretical predictions; for the majority of them, the agreement 
is in fact much better than $2\sigma$, with data basically sitting on 
top of NLO QCD predictions. It is worth noting that in the case of $b$
production at HERA NLO and FONLL predictions~\cite{Cacciari:2001td}
coincide, since the transverse momenta probed are never too large.
This also implies that the treatment of the fragmentation mechanism
is not as delicate as in the case of the Tevatron measurements.

In the case of HERA, the breakthrough that occurred in the last couple of 
years has been mainly due to the fact that, thanks to a much larger
statistics, cross sections could be presented in the experimentally
visible regions (rather than in the form of total rates), and compared
to theoretical predictions obtained by applying the same cuts.
Former experimental results always involved huge extrapolations from
the very narrow visible regions to the whole phase space, performed
with standard parton shower Monte Carlo's, which {\em cannot} give 
sensible predictions for small $\pt$'s (see ref.~\cite{Frixione:2003ei}
for a discussion on this point), a region which gives the dominant
contribution to the total rate. MC@NLO is reliable at small $\pt$'s,
but it is not yet available for photoproduction and DIS processes,
and in any case the publication by the experimental collaborations
of the visible cross sections is always the option to be preferred.

In view of the lesson learned at the Tevatron and HERA, it is 
unfortunate that the measurements of the $\gamma\gamma\to b\bb+X$
cross sections~\cite{Acciarri:2000kd,DObb} suffer from the drawbacks
that prevented a fair comparison between theory and data in $p\bar{p}$
and $ep$ collisions. The three measurements rely on huge extrapolations
from the visible regions to the whole phase space, done with standard
parton shower Monte Carlo's; the uncertainties associated with the
theoretical predictions are too small; the techniques used are
identical. For these reasons, I find it difficult to take at face value
the discrepancies quoted (data are more than a factor of three larger 
than the default NLO predictions), since a careful computation of
all the uncertainties involved (for example, those relevant to the
extrapolation to the whole phase space, which needs to be assessed
by using at least two different Monte Carlos, and ideally the NLO
computations themselves) would presumably show that data lie within
less than $3\sigma$ from theory. Clearly, this statement cannot be 
proved (or disproved) but by the experimental collaborations, which
should follow the strategy set by H1 and ZEUS, of quoting results
relevant to the visible regions. Let me conclude by mentioning that
I'm not aware of any beyond-the-SM mechanism, let alone higher-order
QCD corrections, which could explain these huge discrepancies.

In summary, thanks to improvements on both the experimental and
the theoretical sides, $b$ data at colliders seem to be in fair 
agreement with QCD expectations. I stress that the theoretical 
predictions are essentially based on the NLO computations
of the late 80's~\cite{Nason:1987xz,Nason:1989zy,Beenakker:1990ma},
and that the changes in the predictions of single-inclusive $\pt$ 
spectra are due to a better understanding of the fragmentation
mechanism and to the use of more precisely determined PDF sets.
Newly developed tools, such as MC@NLO, will serve to pin down
discrepancies between theory and data in yet unexplored corners
of the $b\bb$ phase space.

\end{document}